\documentclass[aps,prb,twocolumn,showpacs,superscriptaddress]{revtex4}
\usepackage{epsfig}
\usepackage{times}
\usepackage{amsmath}
\begin{document}

\title{The dynamics of laser droplet generation}

\author{Bla{\v z} Krese}
\email[Electronic address (corresponding author): ]{blaz.krese@fs.uni-lj.si}
\affiliation{Laboratory of Synergetics, Faculty of Mechanical Engineering, University of Ljubljana, A{\v s}ker{\v c}eva cesta 6, SI-1000 Ljubljana, Slovenia}
\author{Matja{\v z} Perc}
\email{matjaz.perc@uni-mb.si}
\homepage[Homepage: ]{http://www.matjazperc.com/}
\affiliation{Department of Physics, Faculty of Natural Sciences and Mathematics, University of \\ Maribor, Koro{\v s}ka cesta 160, SI-2000 Maribor, Slovenia}
\author{Edvard Govekar}
\email{edvard.govekar@fs.uni-lj.si}
\affiliation{Laboratory of Synergetics, Faculty of Mechanical Engineering, University of Ljubljana, A{\v s}ker{\v c}eva cesta 6, SI-1000 Ljubljana, Slovenia}

\begin{abstract}
We propose an experimental setup allowing for the characterization of laser droplet generation in terms of the underlying dynamics, primarily showing that the latter is deterministically chaotic by means of nonlinear time series analysis methods. In particular, we use a laser pulse to melt the end of a properly fed vertically placed metal wire. Due to the interplay of surface tension, gravity force and light-metal interaction, undulating pendant droplets are formed at the molten end, which eventually completely detach from the wire as a consequence of their increasing mass. We capture the dynamics of this process by employing a high-speed infrared camera, thereby indirectly measuring the temperature of the wire end and the pendant droplets. The time series is subsequently generated as the mean value over the pixel intensity of every infrared snapshot. Finally, we employ methods of nonlinear time series analysis to reconstruct the phase space from the observed variable and test it against determinism and stationarity. After establishing that the observed laser droplet generation is a deterministic and dynamically stationary process, we calculate the spectra of Lyapunov exponents. We obtain a positive largest Lyapunov exponent and a negative divergence, \textit{i.e.}, sum of all the exponents, thus indicating that the observed dynamics is deterministically chaotic with an attractor as solution in the phase space. In addition to characterizing the dynamics of laser droplet generation, we outline industrial applications of the process and point out the significance of our findings for future attempts at mathematical modeling.
\end{abstract}

\pacs{05.45.Ac, 05.45.Tp, 42.62.Cf}

\maketitle

\textbf{The dripping faucet is one of the paradigmatic examples of deterministic chaos. Due to the inherently nonlinear interplay between the surface tension, mass of the droplets and the dripping rate, the system exhibits extreme richness of dynamics, culminating in the emergence of deterministically chaotic behavior. Here we propose and examine the dynamics of a conceptually closely related process, which however, is governed by additional physical phenomena. While the complexity of laser droplet generation also relies on the interplay between the surface tension and the increasing droplet mass, the addition of light-metal interaction and the fact that molten metal has different properties than water warrant diversity if compared to the traditional dripping faucet experiment. It is therefore all the more fascinating that from the viewpoint of dynamics, the two processes share deterministic chaos as an inseparable ingredient. Indeed, our analysis reveals that the phase space, reconstructed from the indirect temperature measurements of the metal droplets, is characterized by an attractor having negative divergence and a positive largest Lyapunov exponent. As such, it has all the properties that are characteristic for deterministically chaotic systems. These essential insights into the dynamics of laser droplet generation from experimental data are paramount for the proper introduction of the process to actual industrial applications, as well as to modeling attempts that may further facilitate its understanding. Demonstrating the emergence of chaos in a realistic engineering setup adds to the evergreen nature of the subject, and in this sense, we hope that the study will be inspirational and spawn further research aimed at unraveling the dynamics of laser droplet generation and revealing its full potentials.}

\section{Introduction}

Nonlinear dynamical systems \cite{booksnd} offer a gateway to fascinating phenomena that imbue many facets of our existence. Although frequently going by unnoticed, deterministic chaos, \cite{eckmannXrmp} fractal structures, \cite{aguirreXrmp} synchronization, \cite{snychro} and even the stochastic resonance, \cite{SR} are phenomena that are at the very heart of numerous manmade and natural systems. Be it only a thought or a heart beat, \cite{heartmind} the organization of traffic, \cite{traffic} or the weather forecast, \cite{lorenzXjas} nonlinear dynamics plays an important role in it all. However, while the dripping faucet \cite{faucet} or the flapping of butterfly wings \cite{wings} are paradigmatic examples of deterministic chaos, their omnipresence and universal appeal are in stark contrast with many of the processes in engineering and technical sciences, where the complexity underlying them frequently remains unexplored or at least unknown to the wider audience. Since mathematical models for complex processes are difficult to construct, and are therefore either non-existent or capture only the essential ingredients of the dynamics, one of the common obstacles to overcome is the characterization of the process from experimental or observed data sets.

Nonlinear time series analysis \cite{books} enables the determination of characteristic quantities, for example the number of active degrees of freedom or invariants such as the Lyapunov exponents, \cite{eckmannXrmp} of a particular system by analyzing the time course of one of its measured variables. In the past two decades numerous successful applications of nonlinear time series analysis on data sets from the most diverse fields of research have been reported, \cite{tsarev} and there are still new approaches being proposed to date. One of the most recent advances is the merging of concepts from the theory of complex networks \cite{networks} and time series analysis, \cite{timetonets} particularly also recurrence plots, \cite{recurplots} which together gave rise to new quantifiers for experimental data sets. \cite{advance} Despite the availability of comprehensive and well-documented programs \cite{tisean}, however, there are still branches of science in which the application of these methods could lead to substantial advancements. The failure of applying them is in part certainly due to the difficulties associated with successfully bringing together scientists that are working on very different and seemingly completely disjoint subjects, but also due to the fact that not all data sets are equally amenable to methods of nonlinear time series analysis. While the latter indeed offer tools that bridge the gap between experimentally observed irregular behavior and the theory of dynamical systems, it should be emphasized that this is true foremost if the series under study has properties that are typical of deterministic dynamical systems. \cite{kaplanXprl, schreiberXchaos, brockerXchaos} Moreover, is has to be verified if the observed irregular behavior originated from a stationary system, \cite{schreiberXprl, heggerXprl} for it may solely be a consequence of varying system parameters during data acquisition. These are important issues that have to be addressed, especially on experimental recordings, as we will try to emphasize throughout this work.

In this paper we propose an experiment that allows us to characterize the process of dripping via laser-induced heating of the end of a properly fed metal wire, \textit{i.e.}, the laser droplet generation. Particularly, we are interested in the dynamical properties of this process, which via the analogy with the dripping faucet \cite{faucet} promise to be very interesting. However, although being conceptually similar, we note that the laser droplet generation is governed by additional physical phenomena. While the surface tension and droplet mass also play a crucial role, the additional effects brought about by light-metal interaction, heating, phase transitions and the fact that molten metal has different properties than water distinguish the process significantly from the traditional dripping faucet experiment. In our case a laser pulse is used to melt the end of a vertically placed metal wire. Due to the interplay between the surface tension and the gravity force a pendant droplet is formed from the molten end. The pendant droplet eventually becomes fully detached due to its growing mass and the laser light-matter interaction. During the process of droplet formation the wire has to be properly fed, thereby mimicking the flow rate of water in the dripping faucet experiment. The most important variable to monitor during the process is the temperature of the wire end and the pendant droplet, which we realize indirectly by means of a high-speed infrared camera. Variations in the temperature over time are subsequently obtained as the mean value over the pixel intensity of every infrared snapshot. We start unraveling the dynamics of the observed laser droplet generation via the embedding theorem, \cite{embedding} which enables the reconstruction of the phase space from a single observed variable, thereby laying foundations for further analysis. We use the mutual information \cite{fraseXpra} and the false nearest neighbor method \cite{kennelXpra, heggerXpre} to obtain optimal embedding parameters for the phase space reconstruction. Subsequently, we apply a determinism \cite{kaplanXprl} and a stationarity \cite{schreiberXprl} test to verify that the observed behavior is indeed a consequence of deterministic dynamics and that all the parameters were held constant during data acquisition. After establishing that the studied temperature recording originates from a deterministic and stationary laser droplet generation, we calculate the spectra of Lyapunov exponents, \cite{lyapspec, altlyapspec} whereby a positive largest Lyapunov exponent \cite{lyapmax} and a negative divergence both point towards the fact that the observed dynamics is deterministically chaotic with an attractor as solution in the phase space. We also outline potential industrial applications of the process and give pointers towards its appropriate mathematical modeling. It is notable that the number of droplet-based technologies has increased substantially in recent years. \cite{appli} From this point of view laser droplet generation has an industrial potential especially in joining, \cite{join, join2} where accurate mathematical models would be paramount for further application developments.

The paper is structured as follows. Section II is devoted to the accurate description of the experimental setup and the acquisition of the time series. Section III features results of nonlinear time series analysis, while in the last Section we summarize the paper and outline potential implications of our findings.

\section{Experimental setup}

\begin{figure}
\centerline{\epsfig{file=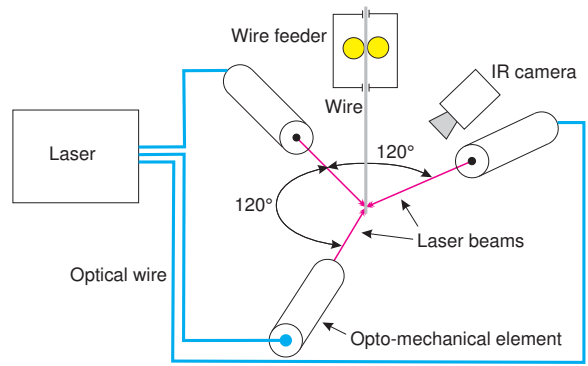,width=8.0cm}}
\caption{Schematic presentation of the experimental setup. The temperature is measured indirectly by means of a high-speed infrared (IR) camera (see main text for details).}
\label{exp}
\end{figure}

The laser droplet generation phenomenologically consists of two phases. In the first phase a laser pulse is used as a source of energy to melt the end of a vertically placed metal wire. From the molten end a pendant droplet is formed due to the action of surface tension and gravity force. Because the surface tension drags the pendant droplet up the wire the latter has to be properly fed. The second phase encompasses the detachment of the pendant droplet from the tip of the wire. To achieve this the surface tension force needs to be overcome, which in our case happens as a result of droplet mass growth. This scenario can be referred to as spontaneous dripping. Notably, a metal droplet can be used in different manufacturing applications. The most promising one is droplet joining, where a molten droplet is placed onto the joining spot. \cite{join, join2} The heat content of a droplet is sufficient to produce a high-temperature weld, whereas the volume of the droplet can be used to fill gaps or bridge dimensional tolerances. Other potential applications include the generation of 3D structures by means of a selective deposition of droplets into layers as well as micro casting. The most common and important process underlying these technologies is laser droplet generation. However, for an effective optimization and control of the process it is essential to know its dynamics. We aim to determine this from experimental data.

In order to study the dynamics of laser droplet generation we have developed an experimental system that is schematically depicted in Fig.~\ref{exp}. The main parts of the experimental setup are the Nd:YAG pulse laser, the opto-mechanical elements, the wire feeder, and the infrared camera. The Nd:YAG laser is used for generating laser pulses with a wavelength of $1.06 {\rm \mu m}$. The maximal laser pulse power is $8 {\rm kW}$ and the pulse duration is between $0.3 {\rm ms}$ and $20 {\rm ms}$. The maximal pulse repetition rate is $300 {\rm Hz}$ with an average power of $0.25 {\rm kW}$. To assure uniform heating of the wire and process symmetry, the laser light is divided into three equal laser beams. By means of the opto-mechanical elements the beams are distributed equiangular along the wire circumference and perpendicularly focused onto the wire's surface. The wire is vertically fed via a controlled wire feeder. Maximal acceleration and velocity of the wire are $20 {\rm m/s^2}$ and $0.3 {\rm m/s}$, respectively. The wire controller is also applied to synchronize the triggering of the laser pulses with the stepwise wire feeding. Since the temperature is the most important variable of the process it was indirectly measured by means of a high-speed infrared camera. Given the properties of the light emitted at the wire end and the pendant droplets, the snapshots were acquired at wavelengths between $3.5 {\rm \mu m}$ and $5 {\rm \mu m}$.

\begin{figure}
\centerline{\epsfig{file=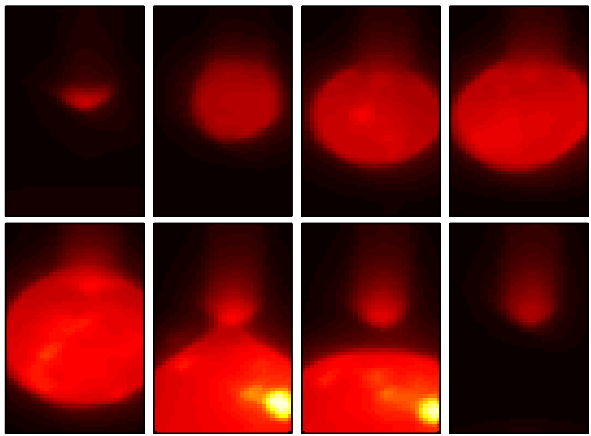,width=8.0cm}}
\caption{A sequence of snapshots from the infrared camera showing growth of the pendant droplet and its detachment. The time increases from the upper left to the lower right panel. We also provide a supplementary video file from the infrared camera, showing the process of laser droplet generation in continuous time (see \textit{snapmovie.avi}).}
\label{snaps}
\end{figure}

\begin{figure}
\centerline{\epsfig{file=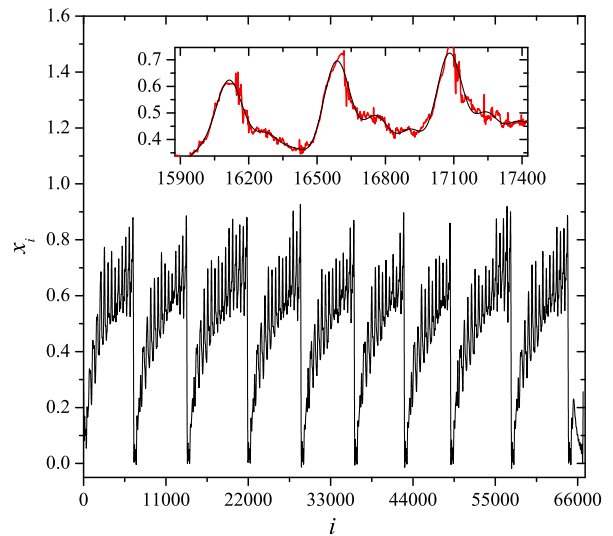,width=8.0cm}}
\caption{The time series, capturing the process of laser droplet generation via the pixel intensity of high-speed infrared snapshots (see Fig.~\ref{snaps}). The main panel shows the time series $x_{i}$ rescaled to the unit interval with an integer time scale. The inset shows the series before (red line) and after (black line) Wiener filtering, removing the high-frequency noisy component that is due to the infrared imaging. No additional noise filtering has been made prior to further analysis.}
\label{series}
\end{figure}

Based on the description of the experimental setup it is obvious that there are a number of parameters that can influence the process of laser droplet generation. For a selected wire material, however, the most important ones are the laser pulse and wire feeding parameters. Proper estimates for the laser pulse parameters can be obtained based on the analysis of the heat balance of a molten pendant droplet. \cite{join2} Parameters of the wire feeding generally depend on the dynamics of the droplet. However, an average wire feed velocity can be estimated based on the desired droplet volume. \cite{join2} Here we have used a nickel wire having diameter $0.6 {\rm mm}$. Other parameters and setup details are as follows: A rectangular laser pulse with power $1440 {\rm W}$, duration $12 {\rm ms}$ and frequency $3 {\rm Hz}$ was used. During the laser pulse, the wire was fed by a triangular velocity profile with a maximal velocity of $0.3 {\rm m/s}$. The motivation behind the usage of a triangular profile of the feeding velocity is two-fold. First, it is simple and transparent enough to be easily implement experimentally, and second, due to its simplicity, it is as non-invasive on the inherent dynamics of the droplet generation as possible, in particular, allowing swift adjustments in accordance with the droplet growth and subsequent detachment. The sampling frequency of the infrared camera was $1428 {\rm Hz}$ at snapshot size of $U \times V = 32 \times 64$ pixels. A short sequence of snapshots is shown in Fig.~\ref{snaps}, where an example of droplet growth and subsequent detachment is depicted (see also the supplementary video \textit{snapmovie.avi}). Finally, the spatiotemporal temperature field was converted into a single scalar time series by calculating the mean value of the pixel intensity of every snapshot according to:
\begin{equation}
x_i = \frac{1}{UV} \sum_{u = 1}^{U}\sum_{v = 1}^{V} x_i^{u,v}.
\label{series_m}
\end{equation}
In Eq.~(\ref{series_m}) $x_i^{u,v}$ is the $i$-th snapshot value of the $(u,v)$ pixel, and thus is as a proxy of the local temperature $T(u,v,t)=x_i^{u,v}$, where $t=i{\rm d}t$ [see also Eq.~(\ref{embed})]. The resulting time series is, rescaled to the unit interval and de-noised by means of a Wiener filter, shown in Fig.~\ref{series}. Upon visual inspection of the time series, lower and higher frequency oscillations can be inferred, which can be linked nicely with the two-phase process of laser droplet generation. Namely, the lower frequency oscillations correspond to droplet volume (mass) and temperature growth, which is followed by a sudden drop of the signal amplitude due to the droplet detachment.

\section{Time series analysis}

We start the time series analysis by applying the embedding theorem, \cite{embedding} which states that for a large enough embedding dimension $m$ the delay vectors
\begin{equation}
{\rm \bf z}(i)=[x_i, x_{i+\tau}, x_{i+2 \tau}, \dots, x_{i+(m-1)\tau}]
\label{embed}
\end{equation}
yield a phase space that has exactly the same properties as the one formed by the original variables of the system. In Eq.~(\ref{embed}) variables $x_i$, $x_{i+\tau}$, $x_{i+2 \tau}$,$\dots$, $x_{i+(m-1)\tau}$ denote values (rescaled to the unit interval for simplicity) of the indirectly measured temperature at times $t=i{\rm d}t$, $t=(i+\tau){\rm d}t$, $t=(i+2\tau){\rm d}t$,$\dots$, $t=[i+(m-1)\tau]{\rm d}t$, respectively, whereby $\tau$ is the embedding delay and ${\rm d}t$ is the sampling time of data points equaling $7 \cdot 10^{-4}{\rm s}$. Altogether, the examined time series consists of $i=1,2, \ldots, 66732$ data points.

\begin{figure}
\centerline{\epsfig{file=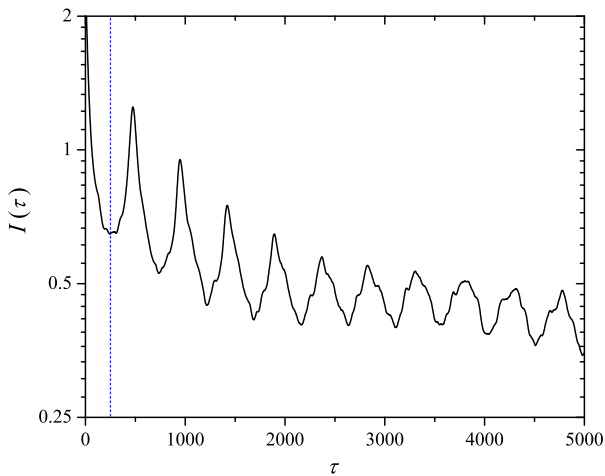,width=8.0cm}}
\caption{Determination of the proper embedding delay via the mutual information method. The first minimum occurs at $\tau \approx 250$ (blue dashed line), which we will use in all subsequent calculations.}
\label{mutual}
\end{figure}

\begin{figure}
\centerline{\epsfig{file=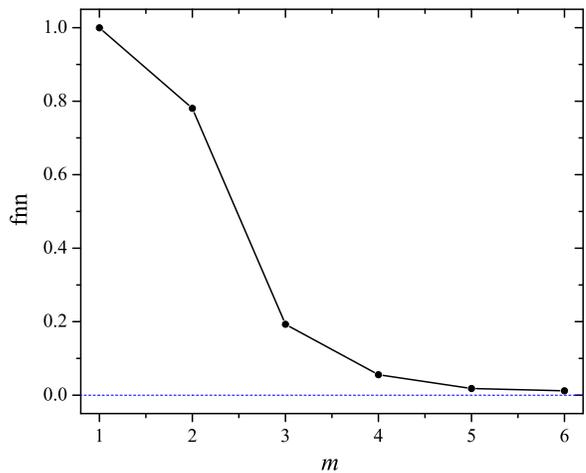,width=7.7cm}}
\caption{Determination of the minimally required embedding dimension. The fraction of false nearest neighbors (fnn) drops close ($<0.01$) to zero (blue dashed line) at $m=5$, which we will use in all subsequent calculations.}
\label{fnn}
\end{figure}

While the implementation of Eq.~(\ref{embed}) is straightforward, we first have to determine proper values for the embedding parameters $m$ and $\tau$. For this purpose, the mutual information \cite{fraseXpra} and the false nearest neighbor method \cite{kennelXpra} can be used, respectively. Since the mutual information between $x_i$ and $x_{i+\tau}$ quantifies the amount of information we have about the state $x_{i+\tau}$ presuming we know $x_i$, \cite{shaw} Fraser and Swinney \cite{fraseXpra} proposed to use the first minimum of the mutual information as the optimal embedding delay. Results presented in Fig.~\ref{mutual} show that the mutual information $I(\tau)$ has the first minimum at $\tau \approx 250$. The false nearest neighbor method, on the other hand, relies on the assumption that the phase space of a deterministic system folds and unfolds smoothly with no sudden irregularities appearing in its structure. By exploiting this assumption one comes to the conclusion that points that are close in the reconstructed embedding space have to stay sufficiently close also during forward iteration. If a phase space point has a close neighbor that does not fulfil this criterion it is marked as having a false nearest neighbor. As soon as $m$ is chosen sufficiently large, the projection effects due to a mapping of the time series onto a space with too few degrees of freedom should disappear, and with them the fraction of points that have a false nearest neighbor (fnn) should converge to zero. \cite{kennelXpra} Note that the method implicitly assumes that a deterministic time series is given as input. This, however, cannot be taken for granted, and indeed a simple extension of the originally proposed false nearest neighbor method \cite{heggerXpre} can be used also as a determinism test. Here we employ the classical algorithm proposed by Kennel \textit{et al.} \cite{kennelXpra} and use the determinism test due to Kaplan and Glass \cite{kaplanXprl}. Results of the false nearest method are presented in Fig.~\ref{fnn}, showing that fnn $\rightarrow 0$ at $m=5$. We will thus use $\tau = 250$ and $m=5$ as input for Eq.~(\ref{embed}) in what follows.

\begin{figure}
\centerline{\epsfig{file=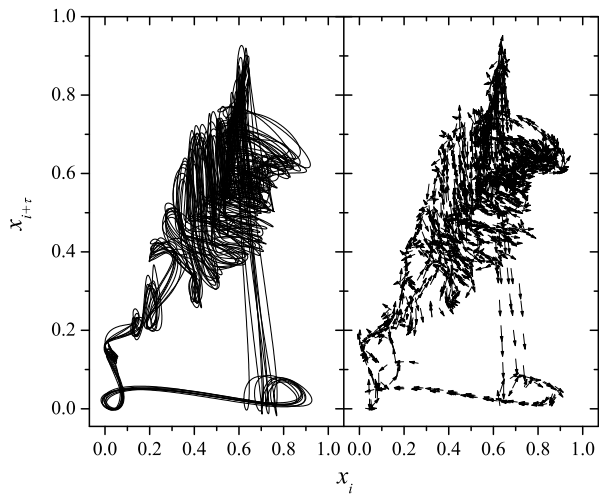,width=8.0cm}}
\caption{Determinism test. Left panel features the reconstructed phase space using $\tau =250$ and $m=5$, while the right panel shows the pertaining approximated directional vector field. Determinism factor of the phase space according to Kaplan and Glass \cite{kaplanXprl} is $\kappa =0.9$.}
\label{determinism}
\end{figure}

Having all the parameters at hand for reconstructing the phase space from the observed variable (see left panel of Fig.~\ref{determinism}), we can proceed by employing the determinism test proposed by Kaplan and Glass. \cite{kaplanXprl} The test is simple but effective, measuring average directional vectors in a coarse-grained embedding space. The idea is that neighboring trajectories in a small portion of the embedding space should all point in the same direction, thus assuring uniqueness of solutions in the phase space, which is the hallmark of determinism. To perform the test, the embedding space has to be coarse grained into equally sized boxes. The average directional vector pertaining to a particular box is then obtained as follows. Each pass $p$ of the trajectory through the $k$-th box generates a unit vector ${\rm \bf e}_{p}$, whose direction is determined by the phase space point where the trajectory first enters the box and the phase space point where the trajectory leaves the box. The average directional vector ${\rm \bf V}_{k}$ through the $k$-th box is then
\begin{equation}
{\rm \bf V}_{k}=n^{-1}\sum_{p=1}^n {\rm \bf e}_{p}
\label{vf}
\end{equation}
where $n$ is the number of all passes through the $k$-th box. Completing this task for all occupied boxes gives us a directional approximation for the vector field. If the time series originates from a deterministic system, and the coarse grained partitioning is fine enough, the obtained directional vector field ${\rm \bf V}_{k}$ should consist solely of vectors that have unit length. Hence, if the system is deterministic, the average length of all the directional vectors $\kappa$ will be close to one. The determinism factor pertaining to the five-dimensional embedding space presented in Fig.~\ref{determinism} that was coarse grained into a $12 \times 12 \times \ldots \times 12$ grid is $\kappa=0.9$, which confirms the deterministic nature of the studied time series. The two-dimensional projection of the directional vector field is shown in the right panel of Fig.~\ref{determinism}. It is also informing to generate surrogates \cite{surrogates} from the studied series, for example by employing the iterative procedure proposed by Schreiber and Schmitz, \cite{schsurro} albeit with the cautionary note that surrogates enable only the rejection (or acceptance) of a given null hypothesis. As such, they do not allow more far-reaching conclusions on the role of stochasticity in the examined series, and thus cannot be used as a substitute for the preceding determinism test. In our case the determinism factor drops to $0.48 \leq \kappa \leq 0.61$ (based on $20$ generated surrogates), which thus rejects the null hypothesis that the laser droplet generation is a stationary Gaussian linear process that has been distorted by a monotonic, instantaneous, time-independent nonlinear function.

\begin{figure}
\centerline{\epsfig{file=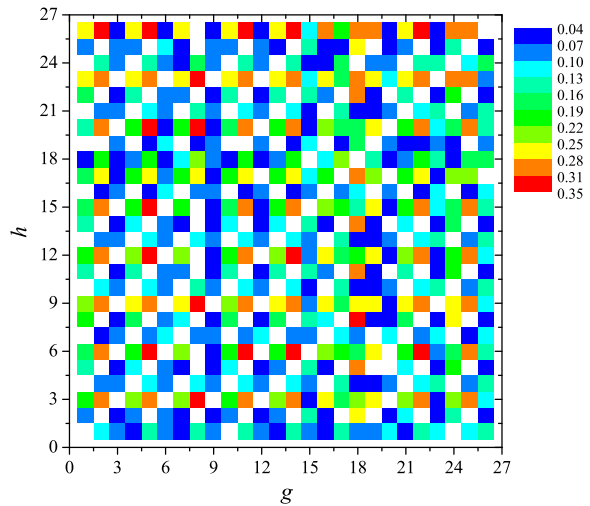,width=8.0cm}}
\caption{Stationarity test. The whole time series was partitioned into $26$ non-overlapping segments each occupying 2500 data points. The color map displays average cross-prediction errors $\delta_{gh}$ in dependence on different segment combinations.}
\label{stationarity}
\end{figure}

It remains of interest to verify if the laser droplet generation is a stationary process. To this purpose, we apply the stationarity test proposed by Schreiber. \cite{schreiberXprl} In general, stationarity violations manifest so that various non-overlapping segments of the time series have different dynamical properties. Since linear statistics, such as the mean or standard data deviation, usually do not posses enough discrimination power when analyzing irregular signals, nonlinear statistics have to be applied. One of the most effective has proven to be the cross-prediction error statistic. The idea is to split the time series into several short non-overlapping segments, and then use a particular data segment to make predictions in another data segment. By calculating the average prediction error $\delta_{gh}$ when considering points in segment $g$ to make predictions in segment $h$, we obtain a very sensitive statistics capable of detecting minute changes in dynamics, and thus a very powerful probe for stationarity. \cite{schreiberXprl} Results presented in Fig.~\ref{stationarity} were obtained by dividing the whole time series into $26$ non-overlapping segments of $2500$ points, thus yielding $26^2$ combinations to evaluate $\delta_{gh}$. Since the cross-prediction errors are uniformly spread across the whole $g-h$ plane, \textit{i.e.}, none of the segments is an exceptionally bad (or good) source of data to make predictions in the other segments, we can refute non-stationarity in the proposed laser droplet generation experiment. An interesting feature of the stationarity test presented in Fig.~\ref{stationarity} is also the emergence of white diagonals, appearing parallel to the main white diagonal. While the latter is expected because there $g=h$ (the segment used for making predictions is also the one we test them against), the other white diagonals appear due to the pseudo-periodicity, which can be inferred from the outlay of the time series (see Fig.~\ref{series}). In particular, although the growth and subsequent detachment of droplets is obviously not a strictly periodic process, the similarity of the individual periods of droplet growth, as well as the similar lengths of time spans between consecutive droplet detachments, nevertheless lead to the recurrent emergence of white diagonals also beyond $g=h$. This, in turn, can also be seen as evidence supporting the determinism in the examined time series.

\begin{figure}
\centerline{\epsfig{file=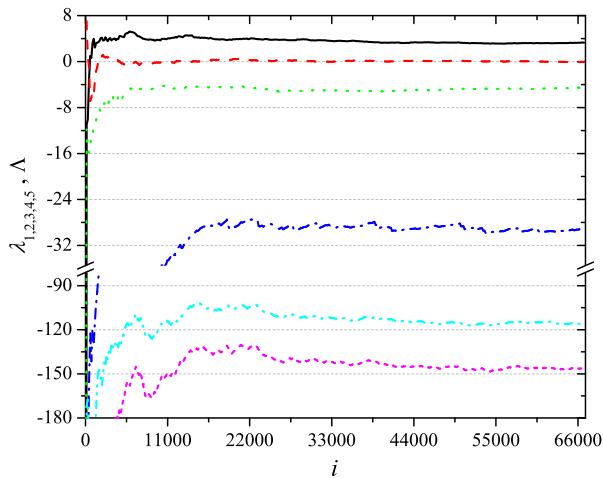,width=8.0cm}}
\caption{Spectra of Lyapunov exponents determined using radial basis functions for the approximation of the flow. From top to bottom the lines depict the convergence of the largest ($\lambda_{1}$) to the smallest ($\lambda_{5}$; most negative) Lyapunov exponent as a function of the discrete time $i$. The lowest line (pink dashed) shows the sum of all five exponents, \textit{i.e.}, the divergence $\Lambda=\sum_{j=1..m}\lambda_{j}$. A linear fit towards the end of the curves gives $\lambda_{1} = (3.2 \pm 0.1){\rm s^{-1}}$, $\lambda_{2} = (0.0 \pm 0.1){\rm s^{-1}}$ and $\Lambda = -(145 \pm 3){\rm s^{-1}}$. Note that the vertical axis has a break.}
\label{lyapspec}
\end{figure}

Finally, we calculate the spectra of Lyapunov exponents $\lambda_j$ where $j=1, 2, \ldots, m$, knowing with reasonable certainty that the obtained results are due to deterministic nonlinear dynamics rather than noise or varying systems parameters during data acquisition. We employ radial basis functions for the approximation of the flow in the phase space. Using the phase space reconstruction parameters obtained above, $M=10$ nearest neighbors of each ${\rm \bf z}(i)$ to make the fit, and the stiffness parameter $r=7$, \cite{lyapspec} the exponents change their sign upon time reversal of the flow and converge robustly as the number of iterations increases. Figure~\ref{lyapspec} features the individual convergence curves, from which we obtain $\lambda_{1} = (3.2 \pm 0.1){\rm s^{-1}}$, $\lambda_{2} = (0.0 \pm 0.1){\rm s^{-1}}$, and the divergence as the sum over all $\lambda_j$ equal to $\Lambda = -(145 \pm 3){\rm s^{-1}}$. From the positive largest Lyapunov exponent, the vanishing second Lyapunov exponent, and the negative divergence, we can conclude that the dynamics of laser droplet generation is deterministically chaotic, and that there exists a stable attractor in the phase space to which any given cloud of initial condition converges in time.

\section{Summary}

We have proposed an experimental setup with the aim of determining the dynamics of laser droplet generation. Using a high-speed infrared camera, we have indirectly measured the spatiotemporal profile of temperature around the molten end of the wire and the pending droplets. Subsequently, the time series was obtained as the mean value over the pixel intensity of every infrared snapshot, and analyzed systematically with methods of nonlinear time series analysis. After reconstructing the phase space from the observed variable, we have verified that the later has properties that are typical for deterministic and dynamically stationary systems. We have shown that the minimally required embedding dimension is five, which altogether suggests that it would be justified to mathematically model the process of laser droplet generation with no more than five first-order ordinary differential equations. Also, we have determined the whole spectra of Lyapunov exponents by approximating the flow in the phase space with radial basis functions. Our calculations revealed that the largest Lyapunov exponent is positive, the second is zero, while the divergence is negative, thus obtaining strong indicators that the observed dynamics is deterministically chaotic with an attractor as solution in the phase space. Thus, although  the laser droplet generation is governed by additional physical phenomena, including light-metal interaction, heating and phase transitions, the dynamics of the process is similar to the one observed in traditional dripping faucet experiments. In addition, the presented results indicate that nonlinearity is an innate ingredient of laser droplet generation, which should be taken into account in future modeling and controlling attempts. We hope that the study will be of value when striving towards a deeper understanding of the examined process and its integration into outlined industrial applications.

\begin{acknowledgments}
We acknowledge support from the Slovenian Research Agency (Program P2-0241; Grant Z1-2032) and the COST-P21 action Physics of Droplets.
\end{acknowledgments}

\end{document}